\documentclass[12pt,preprint]{aastex}
\usepackage{lscape}

\newcommand{\ltsima}{$\; \buildrel < \over \sim \;$}
\newcommand{\simlt}{\lower.5ex\hbox{\ltsima}} 
\newcommand{\gtsima}{$\; \buildrel > \over \sim \;$}
\newcommand{\simgt}{\lower.5ex\hbox{\gtsima}} 

\renewcommand{\arcsec}{\mbox{$^{\prime\prime}$} }

\begin{document}

\title{X-ray Line Emitting Objects in XMM-Newton 
observations: the tip of the iceberg.}

\author{T. Maccacaro$^{1}$, V. Braito$^{1}$, R. Della Ceca$^1$, P. Severgnini$^{1}$, 
and A. Caccianiga$^{1}$}

\affil{$^1$ INAF - Osservatorio Astronomico di Brera, via Brera 28, 20121
Milan, Italy; tommaso@brera.mi.astro.it, braito@brera.mi.astro.it, rdc@brera.mi.astro.it,
paola@brera.mi.astro.it, caccia@brera.mi.astro.it}

\begin{abstract}
We present preliminary results from a novel search for X-ray Line Emitting
Objects (XLEOs) in XMM-Newton images. Three sources have been detected in a
test-run analysis of 13 XMM-Newton observations. The three objects found are
most likely extremely absorbed AGN characterized by a column density $N_H
\sim 10^{24}$ cm$^{-2}$. Their redshift has been directly determined from the
X-ray data, by interpreting the detected emission line as the 6.4 keV Fe line. The
measured equivalent width of the X-ray line is, in all three cases, several
keV. This pilot study demonstrates the success of our search method and
implies that a large sample of XLEOs can be obtained from the public
XMM-Newton data archive. 
\end{abstract}


\section{Introduction}

Both Active Galactic Nuclei (AGN) and clusters of galaxies are known to
exhibit, under the right circumstances, a prominent emission line in their
spectrum. 

In the case of AGN the strongest line in the 2-10 keV range is the ``neutral"
Fe K${\alpha}$ line at 6.4 keV (Mushotzky, Done and Pounds, 1993). In
``classical" optically Type 1 AGN, where both the continuum and the Fe-emitting
region are viewed directly, the equivalent width (EW) of the Fe line is small
and typically less than 200 eV (e.g. Nandra and Pounds, 1994). On the
contrary, in the case of optically Type 2 AGN the Unification Scheme predicts
that obscuring material (a molecular torus?) near the active nucleus blocks the direct
view to the central engine. In these cases, if the ``line emitting"  material
is exposed to a stronger continuum than the one detected by the observer,
we should observe very large EW (Krolik and Kallman, 1987). For example, according to the modeling 
presented in Levenson et al. (2002), EW up to 10 keV are expected in the case of
the most extreme Compton-thick ($N_H > 10^{24}$ cm$^{-2}$) deeply buried AGN.

Furthermore it is now clear and well established that the classification of
galactic nuclei based only on their optical spectrum provides an incomplete
description of their nature (e.g. Vignali et al., 1999; Della Ceca et al.,
2002; Maiolino et al., 2003) and that X-ray data are fundamental for the
recognition and classification of galactic nuclei (see, among others,
Severgnini et al., 2003 and references therein). For example, studies of active
objects at IR and X-ray wavelengths indicate a concomitant AGN and starburst
activity (Fadda et al., 2002) which seems to happen in a high-density medium
($N_H > 10^{24}$ cm$^{-2}$) characterized by high dust extinction of the UV
optical flux and strong photoelectric absorption of the soft X-rays. For
these objects a detailed investigation around the iron line at 6.4 keV seems
to be at the moment the only way to find evidence of the presence of an AGN
(see e.g. Della Ceca et al., 2002). The existence (and the census) of such
optically elusive AGN may have profound implications on the Cosmic X-ray
Background (CXB) synthesis models as well as on our understanding of the
accretion history vs. the stellar history of the Universe. As already
discussed and emphasized by Levenson et al. (2002), the prominent Fe
K${\alpha}$ line, which seems to be a very common feature of very obscured 
AGN, ``can be exploited to find more of them". 

The second class of extragalactic objects known to exhibit a prominent iron
line (the ``ionized" Fe K${\alpha}$ line at 6.7 keV) in their X-ray
spectrum are clusters of galaxies. X-ray selected cluster surveys in the ROSAT
era, based on the source extent, (e.g. the RDCS survey: Rosati et al., 1995;
1998) have led to routine identification of clusters out to z = 0.85, with
only a few examples at higher redshift (see Rosati, Borgani and Norman, 2002
and references therein). The search proposed here has the potential for
finding clusters of galaxies up to a redshift of ~1.6, thanks to their 6.7 keV
ionized Fe K${\alpha}$ line. Indeed this line is clearly detected in the
X-ray spectra of RDCS1252.9-2927 (Rosati et al., 2004), a massive cluster at z
= 1.24. We are aware, however, that since the EW of the Fe line is smaller in
cluster of galaxies than in absorbed AGN, the search for clusters of galaxies
through their emission lines could be more difficult and/or less efficient 
than the search for heavily absorbed AGN. 

The unprecedented combination of high throughput ($\sim$600 cm$^{2}$ effective
area in the $\sim$2.3$-$5.0 keV energy band) and high energy and spatial
resolution (100 eV FWHM at 3 keV and $\sim$15\arcsec respectively) offered by
the ESA XMM-Newton X-ray telescope (just the one mirror module with the
Epic-pn detector), has convinced us of the possibility of searching the sky 
in a
novel way, to discover weak X-ray sources whose emission is mainly
concentrated in line flux. Without a dedicated detection technique these
sources, X-ray Line Emitting Objects (XLEOs), would be hard or
impossible to find because their line emission is usually diluted within
the typical broad energy interval of the X-ray images (a few keV). Even when
detected also by ``standard" techniques, as is the case of the three sources 
presented in this letter (the X-ray brightest candidates among those found from
the test run analysis of 13 XMM-Newton images), these sources would remain 
indistinguishable (unless very bright) from the other thousands of serendipitous
sources because of their very limited number of photons. None of the usual
diagnostic diagrams (hardness ratios, Fx/Fopt, etc.) have the capability of
clearly separating these sources from the multitude of generic serendipitous sources.
Even an automated standard X-ray spectral analysis of a huge number of 
serendipitous sources will probably fail to identify them since the paucity of
counts will force a rather large binning over the energy axis.

\section{The method of analysis}

The idea is based on the extension to the energy axis of the usual source
detection techniques that are typically restricted to the spatial coordinates.
The selected energy range (2.3$-$5.2 keV) implies that the sources
discovered will have a redshift in the interval 0.3$-$1.7, when we consider 
that
the most prominent X-ray line is typically the iron line. 
The energy range chosen minimizes variations in the effective area (there is a
sharp change below $\sim$ 2 keV, due to the Au M edge). As a consequence no
strong or sudden variations are expected in the background (cosmic +
instrumental + source continuum) across the energy range considered. It will
be interesting at some  point to extend the line search to the largest
possible range so as to sample from ``local" to z $\sim 5$ objects.

A detailed discussion of the
detection algorithm and of its performance will be given in Braito et al.
(2005, in preparation). In the interest of clarity we briefly summarize here the basic
properties of our search. Each XMM-Newton image (data cube), has been 
raster-scanned with a 3-D detection cell of $50\arcsec \times 50\arcsec \times
200$ eV. The
step of the scan is half a cell width, in both the spatial axis and in the
energy axis. At each X,Y position, counts in the energy range 2.1$-$5.4 keV
are considered and binned in 200 eV bins. This range is one bin larger, on
each side, than the range of interest for the line detection. The background
level along the energy axis is then determined, after an iterative process to
remove the possible presence of an excess due to a line, with a quadratic fit
to the data. For the purpose of line detection, the source continuum, if
present, is treated as background. Candidates XLEOs are then flagged if a
particular energy bin in the range 2.3$-$5.2 keV contains a number of counts such
that their probability of being a fluctuation of the estimated background
level is smaller than $\sim 10^{-5}$.

\section{Results from the test-run}

We have test-run the above algorithm on a small number of data sets, thirteen,
in order to assess its success rate and to fine-tune the various critical
parameters (e.g. cell size in both spatial and energy dimensions, background
determination, threshold for XLEO candidate flagging, etc.). Only the Epic-pn
data have been used. We have chosen the XMM-Newton test observations among
those at high galactic latitude, with an exposure time in the range from $\sim
1.5\times 10^4$ s to $4.5\times 10^4$ s. Of the candidates found during the
test-run, we present here the three most significant ones, which are also 
characterized by the highest continuum. The three candidates were flagged
because of the detection of 9 counts in the energy bin 3.7$-$3.9 keV (1.4
counts expected), 7 counts in the bin 4.5$-$4.7 keV (0.7 counts expected) and
10 counts in the bin 4.6$-$4.8 keV (1.2 counts expected) respectively (see
inset in Figure 1a$-$c). 
 
With respect to the statistical significance of the lines, we would like to stress
the following. In the energy bin where the line is detected, the probability that
the excess seen is a random noise fluctuation is $1.6 \times 10^{-5}$, $8.9 \times
10^{-6}$ and $5.8 \times 10^{-7}$ for the three sources respectively. This is
computed as the probability of observing 9, 7 and 10 counts (or more) when 1.4,
0.7, and 1.2 counts are expected. However, one should also consider the number of
energy bins (14) and of the spatial cells searched. Although we have
raster-scanned the whole image (but have ignored cells affected by gaps and
edges), in analyzing the preliminary results of our search we have conservatively
considered here only those XLEO candidates coincident with an X-ray source,
unambiguously detected by other means (i.e. standard XMM Science Analysis System).
Indeed these three sources were independently found by the standard detection
algorithm used to produce the 1XMM source catalog (XMM-Newton SSC,
2003)\footnote{http://xmmssc-www.star.le.ac.uk/newpages/xcat\_public.html} and are
listed there. Thus, given that there are $\sim 600$ sources in the set of images
used at the brightness level considered, the
number of trials is $14 \times 600$ and the resulting total number
of false detections expected is $\sim 0.1$. Furthermore, we can adopt the X-ray
positions reported in the 1XMM catalog since they are significantly better than
those produced by our algorithm. 

For all three sources, we did go back to the XMM-Newton data to extract the
broad band spectrum, estimating the background from a nearby, source-free
region. Apertures of 20\arcsec, 20\arcsec and 25\arcsec radius were used to
extract the source counts, and of 23\arcsec (because of the closeness to an
intra-chip gap), 40\arcsec and 50\arcsec radius for the background counts.
When possible, (XLEOJ153241$-$082906 and XLEOJ220425$-$015123) data from the
MOS detectors were also used in order to  maximize the available statistics
and have contributed to the determination of the values given in Table 1. In
the interest of clarity, however, the MOS data have not been plotted in Figure 1.

Inspection of the distribution of the resulting net counts reveals a rather
weak continuum, the possible emission line that prompted the detection, and an
excess of counts at high energies. All three sources seem to be characterized
by an X-ray spectrum consisting of a two component model: a ``leaky absorbed"
power law continuum plus an emission line. This is typical of absorbed AGN and
thus suggests that these three sources are indeed AGN. Although this is not
the first time that an AGN is recognized as such and classified directly from
the X-ray data (e.g. AXJ2254+1146, Della Ceca et al., 2000), to our
knowledge, it is the first time that this happens beyond the local universe.

In Figure 1 (a through c) we report the X-ray spectrum of the three sources and
the distribution of the total counts (inset) in the 3-D cells, which has led
to the line detection. From the position of the emission line, and assuming
that it is due to cold Fe at 6.4 keV, it is possible to derive, directly from
the X-ray data, the redshift of the sources. The values are reported in Table
1 where the basic X-ray properties of these three sources are summarized. We
stress that the values derived should be considered as indicative given the
very low statistics involved (the three sources have of the order of 50 net
counts each in the Epic-pn detector). In the fitting procedure, we did not apply Occam's razor, rather
we have assumed a reasonable model (see above), and determined a set of values
that well describe the data. This is why no formal errors are quoted on the 
derived quantities.

Inspection of deep optical material reveals the presence of 1$-$2 candidates
consistent (on positional ground) with being the optical counterparts of the
X-ray sources. Their magnitudes are in the range 22.0$-$23.5. Considering,
for each X-ray source, the brightest candidate, the resulting
log($F_{x}/F_{opt}$) are in the range $\sim 1.2-1.8$. These values are
typical of absorbed AGN (see, among others, Fiore et al. 2003 and Della Ceca et
al. 2004) and thus further support our proposed
identifications. Optical spectroscopy is of course needed to validate these
results and has been proposed, together with follow up XMM-Newton observations.

\section{Discussion}

The results presented here are very preliminary. We have no doubts about the
reality of the sources and we are confident on the presence of the X-ray
emission line in the source spectra. Three objects are too few to derive
``general properties". Also, we still have to determine the complex
visibility function of our novel algorithm, the volume investigated for a
given line luminosity/EW, the sensitivity of our search as a function of the
spectral characteristics of the sources (continuum slope, intrinsic N$_{H}$, iron
density etc.) and of their redshift. Once this is obtained and a larger sample
is assembled, it will be possible to derive volume densities, population
properties and more general conclusions.

This said, and with the caveat of the low statistics and un-quantified
selection effects, we would like to note a few things: 

1) all three objects found are most likely AGN in, or close to, the
Compton thick regime. They have resulted from the analysis of 13 XMM-Newton
observations. This initial success rate is very encouraging since quite a
large sample can be readily assembled using the hundreds of suitable
XMM-Newton observations  already available from the ESA archive. If follow up
studies confirm that  the three objects found are indeed AGN absorbed by
a column density of $\sim 10^{24}$ cm$^{-2}$ and more, then we have a
formidable tool to find and study these AGN, beyond the local universe and up
to redshifts of cosmological relevance;

2) for the brightest XLEOs, characterized by a detectable continuum, it will
be often possible to recognize the nature of the object and to determine its
redshift directly from the X-ray data, especially in cases like the ones
reported here of absorbed AGN;

3) the redshifts of the three objects found (0.64, 0.45, 0.42) are in the
middle to low part of our window (0.3 $\simlt z \simlt 1.7$). To understand
whether this is primarily due to selection effects or whether it reflects an
intrinsic distribution, requires a detailed mapping of the visibility function
of our algorithm;

4) no cluster has been found in this test run. It is not clear at this point
whether this result is meaningful. A space density of $\sim 15-20$ clusters
per sq. deg. (with z $>$ 0.3) and with an X-ray flux $> 8\times 10^{-15}$ erg
cm$^{-2}$ s$^{-1}$ (0.5$-$4.0 keV) has been reported (Andreon, 2004, private 
communication). It is not obvious, however, how this limiting flux compares
with our detection threshold based on a line contrast, neither is known the
fraction of clusters with a prominent (detectable) emission line. To attack
this issue we plan to run our algorithm on a number of XMM-Newton observations
known to contain high redshift clusters of galaxies.

To conclude, prompted by the power of the XMM-Newton mirror module plus
Epic-pn detector we have developed a novel algorithm to search X-ray datacubes
(X,Y,E) for X-ray Emission Line Objects. A test-run on a dozen XMM-Newton
observations has proven this search extremely successful, yielding the
discovery of three sources, most likely highly absorbed AGN ($N_H > 10^{24}$
cm$^{-2}$), and allowing a crude, but direct, determination of their redshift
from the detected X-ray line. If these findings are confirmed by a more
extensive analysis and by follow up studies of the candidate objects, we have
unlocked the doors to the long sought large sample of highly absorbed, Compton
thick AGN, a key to the Cosmic X-Ray Background synthesis models. Also, it is
possible that other rarer objects will be detected, as the volume searched
increases.

\acknowledgements

The idea of searching for Line Emitting Objects in X-ray images dates back to
1993 when one of us (TM), during a conference by Professor Y. Tanaka, was
impressed by the quality of the X-ray spectra provided by the ASCA SIS, the
first large field of view, solid state, X-ray detector flown. Unfortunately
ASCA lacked the throughput and the angular resolution necessary to make the
search for XLEO feasible and we had to wait for about 10 years to convert the
idea into a successful experiment. We thank S. Andreon, L. Maraschi and A.
Wolter for useful suggestions and stimulating discussions. PS acknowledges
a research fellowship from the Istituto Nazionale di Astrofisica (INAF).

\clearpage


\clearpage
\begin{figure}
\rotatebox{-90}{
\epsscale{0.8}
\plottwo{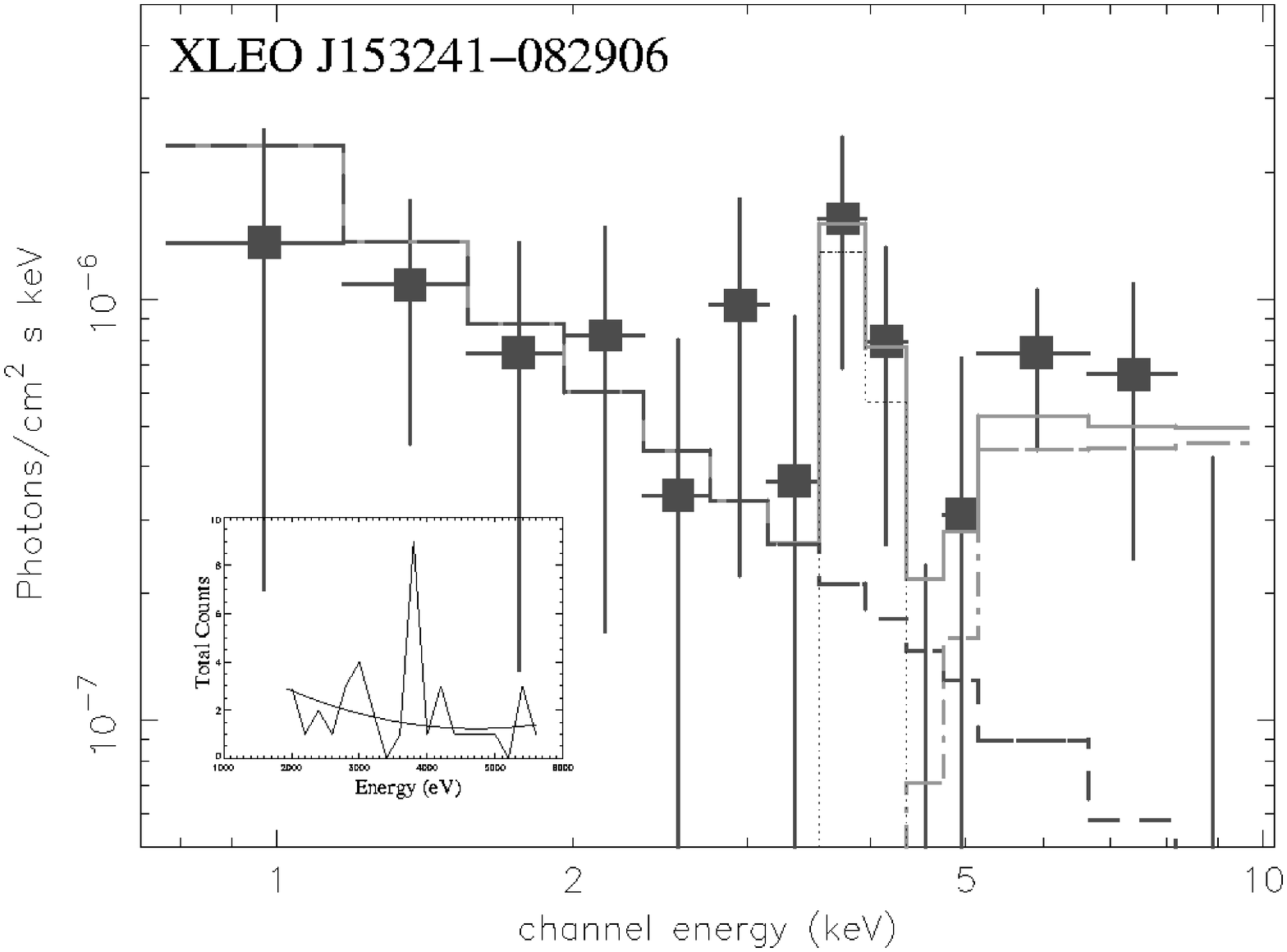}{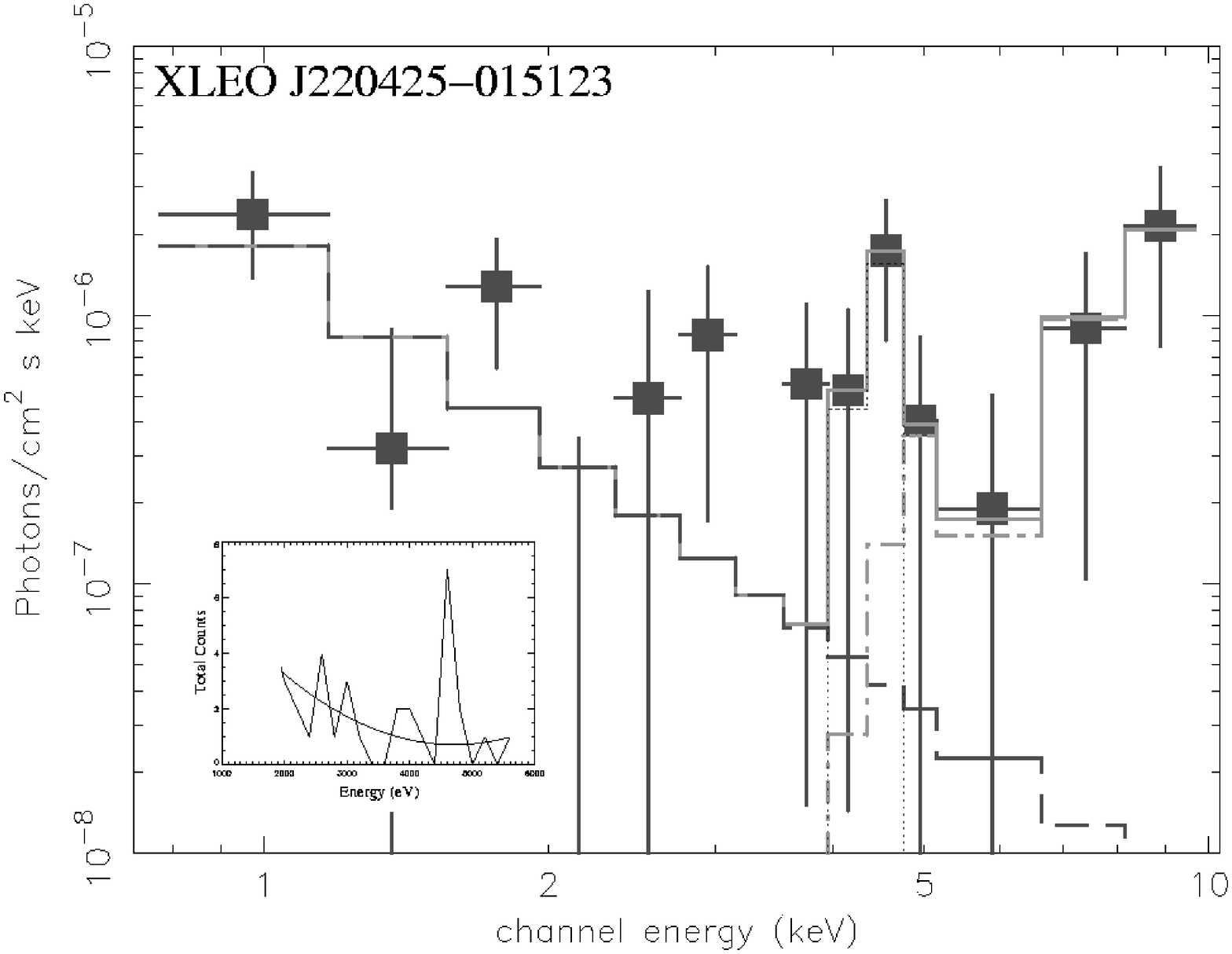}
\epsscale{.4}
\plotone{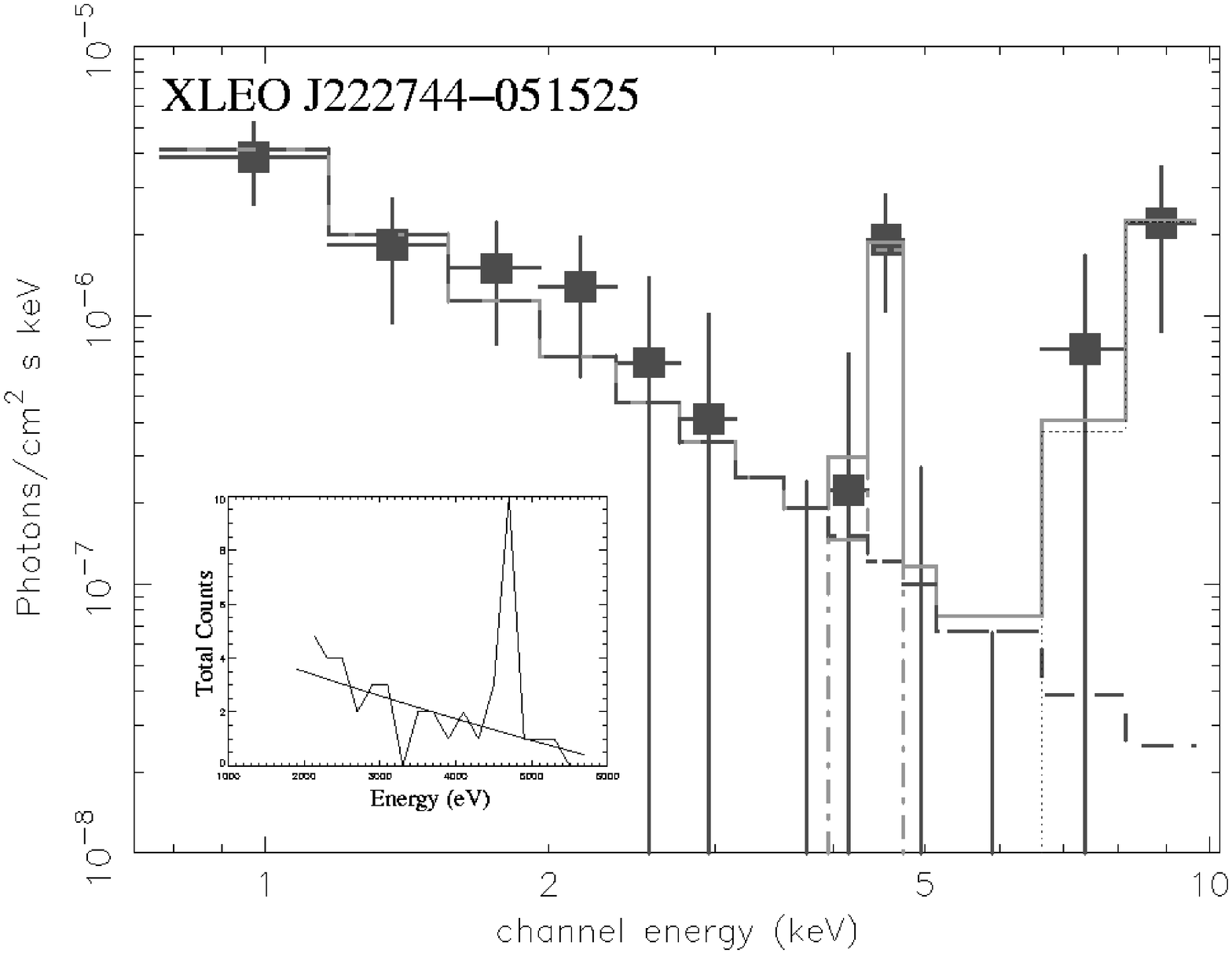}}

\caption{X-ray spectrum, in photon units, of XLEOJ153241$-$082906 (panel a),
XLEOJ220425$-$015123 (panel b) and  XLEOJ222744$-$051525 (panel c), obtained with
XSPEC v. 11.2. The unfolded spectrum (points) and best fit model 
(lines) are shown. For each source, in the inset, the total counts
distribution from the 3-D detection cell raster scan that has led to the line
detection is shown,  together with the estimated background (noise + source
continuum) level. See text for details.} \end{figure}

\clearpage

\begin{table}
\begin{center}
\caption{X-ray data}
\begin{tabular}{lclllllll}
\hline
\hline
Source name          & X-ray position            & E$_{line}$   & EW   & $N_H$             & Flux              & z$_x$   & $\Gamma$  \\
                     &   J2000.0                 &  keV         & keV  &  cm$^{-2}$        & 2$-$10 keV        &         &           \\
(1)                  & (2)                       & (3)          & (4)  & (5)               & (6)               & (7)     & (8)        \\
\hline

XLEOJ153241$-$082906 &	15 32 40.6 $-$08 29 03 &  3.9	      & 5    & $9\times 10^{23}$ & $4\times 10^{-14}$ & 0.64  & 1.9 \\
XLEOJ220425$-$015123 &  22 04 24.9 $-$01 51 28 &  4.4         & 5    & $4\times 10^{24}$ & $8\times 10^{-14}$ & 0.45  & 2.7 \\
XLEOJ222744$-$051525 &  22 27 44.3 $-$05 15 24 &  4.5         & 7    & $8\times 10^{24}$ & $9\times 10^{-14}$ & 0.42  & 2.4 \\
\hline
\hline
\end{tabular}
\end{center}
Column (1): Source name; column (2): X-ray source position; column (3) line energy
position; column (4) observed line equivalent width, computed with respect to the
absorbed component; column (5): intrinsic absorbing column density; column (6):
observed flux in erg cm$^{-2}$ s$^{-1}$; column (7): redshift estimated from the
X-ray line position; column (8): photon index of the power law component.

\end{table}

\newpage





\end{document}